\documentclass{aa}
\usepackage{graphicx,epsfig} 
\begin{document} 
 
\newcommand{\ea}{{et al.}} 
\newcommand{\beq}{\begin{equation}} 
\newcommand{\enq}{\end{equation}} 
\newcommand{\bfg}{\begin{figure}} 
\newcommand{\efg}{\end{figure}} 
\newcommand{\bfa}{\begin{figure*}} 
\newcommand{\efa}{\end{figure*}} 
\newcommand{\bea}{\begin{eqnarray}} 
\newcommand{\ena}{\end{eqnarray}} 
\newcommand{\dd}{{\rm{d}}} 
\newcommand{\dg}{^{\rm{o}}} 
\newcommand{\thetao}{{\theta_{\rm{o}}}} 
\newcommand{\nuo}{{\nu_{\rm{o}}}} 
\newcommand{\thetae}{{\theta_{\rm{e}}}} 
\newcommand{\nue}{{\nu_{\rm{e}}}} 
\newcommand{\re}{{r_{\rm{e}}}} 
\newcommand{\rin}{{r_{\rm{in}}}} 
\newcommand{\rout}{{r_{\rm{out}}}} 
\newcommand{\const}{{\mbox{const}}} 
\newcommand{\bmath}[1]{\mbox{\boldmath{${#1}$}}} 
 
\def\lb#1{{\protect\linebreak[#1]}} 
 
\def\apj{Astroph. J.} 
\def\apjl{Astroph. J. Lett.} 
\def\asa{Astron. \& Astroph.} 
\def\mnras{MNRAS} 
\def\phr{Phys. Rev.} 
\def\phrl{Phys. Rev. Lett.} 
 
\def\ltsima{$\; \buildrel < \over \sim \;$} 
\def\gtsima{$\; \buildrel > \over \sim \;$} 
\def\simlt{\lower.5ex\hbox{\ltsima}} 
\def\simgt{\lower.5ex\hbox{\gtsima}} 
\def\ebf{{\bf e}} 
\def\Ebf{{\bf E}} 
\def\dpar{{\bf \partial}} 
\def\ka{{\rm K}$\alpha$} 
\def\kb{{\rm K}$\beta$} 
\def\ee{\`{e}} 
 
\title{Evidence for a relativistic iron line in GRS 1915+105} 
  
\author{A.\ Martocchia\inst{1}, G.\ Matt\inst{1}, V.\ Karas\inst{2}, 
        T. Belloni\inst{3}, M.\ Feroci\inst{4} } 
\offprints{A.\ Martocchia, {\sf martocchia@fis.uniroma3.it}} 
\institute{ 
Dipartimento di Fisica, Universit\`a degli Studi ``Roma Tre'', 
Via della Vasca Navale 84, I--00146 Roma, Italy 
	\and 
Astronomical Institute, Charles University Prague, 
Faculty of Mathematics and Physics, V Hole\v{s}ovi\v{c}k\'ach 2, 
CZ--180 00 Praha, The Czech Republic 
\and 
Osservatorio Astronomico di Brera, via E. Bianchi 46, I--23807 Merate, Italy 
\and 
IAS/CNR, Area di Ricerca di Tor Vergata, Via Fosso del Cavaliere 100,  
I--00133 Roma, Italy  
}    
\date{Received... Accepted...} 
 
\abstract{We report on the discovery of a relativistic iron
\ka\ fluorescent emission line in the
{\it BeppoSAX} spectrum of the {\it microquasar} GRS 1915+105,
taken on April 19, 1998, when the line was unusually intense. 
The feature is broad and skewed,  
clearly indicating emission from the innermost regions of the 
accretion disc.
The inner emitting orbit is larger  
than the innermost stable orbit, even in 
Schwarzschild metric: thus, a non-zero BH spin is not required
by these data.
 }

\authorrunning{A. Martocchia et al.}
\titlerunning{Evidence for a relativistic iron line in GRS~1915+105}
\maketitle

	\section{Introduction}
	\label{intro}

\subsection{Iron line diagnostics of relativistic accretion discs }

Black Hole (BH) accreting systems are expected to
exhibit spectral features whose form is determined 
by reprocessing of primary high-energy photons
by gas swirling around the central body. Resulting spectral  
distortions are of utmost importance because they probe the effects 
of strong gravity, which are otherwise difficult to reveal.
General-relativity (GR) calculations of predicted spectra, and their  
comparison with observational data, may help to measure physical 
parameters of BHs, such as their masses and angular momenta 
(e.g.\ Martocchia 2000, and references therein). 

The iron \ka\ fluorescent emission is specially 
important as a diagnostic of GR effects (e.g. Fabian et 
al. 1989; Laor 1991; Matt et al. 1992; Martocchia \& 
Matt 1996; Martocchia et al. 2000). In case it is 
originated in the innermost regions of an accretion disc, 
it gives the possibility, in principle, of distinguishing 
between a rotating (Kerr) and a static (Schwarzschild) 
central BH. This is because the resulting line profile is
affected both by Special Relativity (Doppler,
including transverse, shifts) and General Relativity
(energy shift, lensing) effects. It is very broad and 
asymmetric, the exact shape critically depending
on the geometrical parameters of the emitting region
(disc inner and outer radii, inclination angle). For the same
set of parameters, differences between Schwarzschild and 
extreme Kerr line profiles are rather subtle. However, the 
innermost stable orbits of the disc are different in the two
cases - corresponding respectively to 6 and 1.23 gravitational 
radii - and the profiles are very different for the two values. 
The measurement of an innermost emitting radius less than 
6$r_{\rm g}$ can therefore be the signature of a spinning BH.

This diagnostic has been successfully used in the case 
of Seyfert galaxies (e.g. Tanaka \ea, 1995, Wilms \ea, 2001), 
while in galactic BH candidates ionization of the inner 
accretion flow generally prevents fluorescent emission 
from being effective. Nevertheless, intense, broad Fe 
\ka\ emission has been observed in Cyg X-1 with both {\it BeppoSAX} 
(Frontera \ea, 2001) and {\it Chandra} (Miller \ea, 
2002a). Large, possibly relativistic \ka\ profiles have been 
detected also in XTE J1748-288 (Miller \ea, 2001) and XTE J1650-500 
(Miller \ea, 2002b).

\subsection{GRS 1915+105 }

GRS 1915+105 is one of the so-called {\it microquasars}, 
Galactic jet sources with properties similar 
to those of quasars, but on a smaller, stellar scale. 
It was extensively studied since its discovery by the  
WATCH experiment on {\it GRANAT} in 1992 (Castro-Tirado \ea\ 1992).  
Due to very large interstellar absorption, the optical 
counterpart of this source was not easily found: only 
recently, Greiner \ea\ (2001a) reported results of a  
spectroscopic analysis in the H and K bands, which suggest  
that the mass-donating star is a K-M III star, i.e. that 
GRS 1915+105 belongs to the class of low-mass X-ray binaries. 

Greiner \ea\ (2001b) used these infrared  
spectroscopic observations to infer the companion's orbital  
period - which is $\sim 33.5$ days -
and determine the mass function $f(M_{\rm c})\simeq9.5  
M_\odot$. Since the donor's mass, for K-M III stars 
of that kind, is known to be $\sim 1.2 M_\odot$,  
and the inclination is known, too, the mass of the  
central compact object has been constrained to  
$M_{\rm c}=14\pm4 M_\odot$, i.e. well above the 
standard neutron star mass limit. 

Weak variable radio/IR emission connected with jets has  
been often detected (Mirabel \& Rodriguez 1994). 
Measurements of the jets' superluminal motion allowed  
to estimate the source distance $D \sim 12.5$ kpc and  
inclination $i \sim 70^{\rm o}$  
(Rodriguez \& Mirabel, 1999; Fender \ea, 1999).  

GRS 1915+105 is well-known for the extremely wide variety of  
its variability modes, ranging over all possible 
bands and timescales (e.g. Markwardt \ea\ 1999,  
Belloni \ea\ 1997a,b; Greiner \ea\ 1996).  
Time-resolved observations have been intensively  
performed especially with the {\it Rossi X-ray Timing Explorer} 
({\it RXTE}) satellite mission. Belloni \ea\ (2000), for instance,  
managed to classify these observations into 12 separate  
classes based on count rate and color characteristics.  
In this way, the source variability could be explained  
in terms of transitions between three basic states:  
{\it two soft states} with a fully observable disc at two  
different temperatures (``A" with the  
lower temperature, ``B" with the higher), and {\it a hard state}  
corresponding to the non-observability of the innermost parts of  
the accretion disc (state ``C"). State ``C'', with low flux  
and hard color, can last uninterruptedly for weeks or even months. 
The most popular model for the source's large scale variability is 
the one proposed by Belloni \ea\ (1997a,b), who invoked
the onset of a Lightman-Eardley thermal-viscous instability 
which blows-off the innermost, radiation-pressure supported 
part of the (optically thick) disc.  
The hard state (C) would then correspond to a steady replenishment,  
which takes place on a viscous timescale.  
Indeed, evidence of the temporary disappearance and  
subsequent restoring of the inner portion of the  
accretion disc comes from simultaneous {\it BeppoSAX}  
and Ryle radiotelescope observations (Feroci \ea, 1999). 

   \begin{figure}
   \centering
   \includegraphics[width=8.5cm]{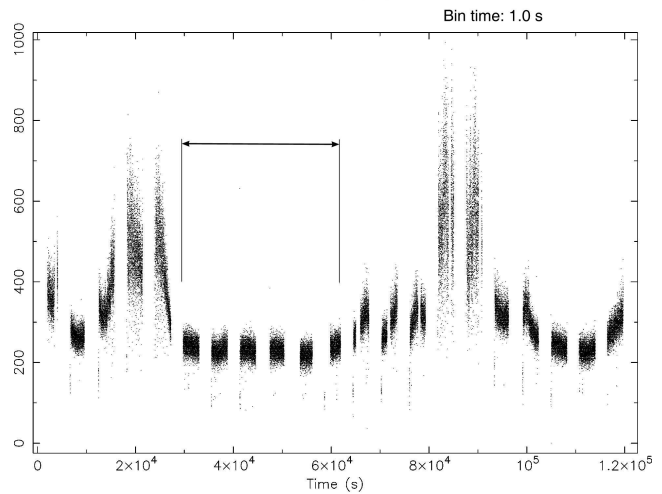}
\caption{The MECS lightcurve of GRS 1915+105 during the April 19, 1998 
{\it BeppoSAX} observation, which lasted $\sim 120$ ks. 
The time intervals we considered are those marked (comprised between 
$t \sim 3\times 10^4$ and $t \sim 6.2\times 10^4$ s). In this paper, the 
fifth of these intervals is discussed in more detail. } 
\label{fig:1915lc}
   \end{figure}

The iron \ka\ emission line in GRS~1915+105 is usually 
not very intense, and is therefore difficult to study in 
detail - not to mention the dramatic spectral  
variability of the source,  
which requires selecting short time intervals,  
so reducing the statistical quality of the spectra.  
Lee \ea\ (2001) recently reported that neutral Fe  
K$\alpha$ emission is likely present in the high-resolution 
data of a low-hard state, collected with {\it Chandra} 
HETGS, together with Fe photoelectric edges and  
absorption lines due to highly ionized matter. These
features indicate that ionized iron, probably belonging
to the accretion disc itself, is present on the line of sight
with abundance larger than solar.

Actually, during the several observational campaigns  
performed by {\it BeppoSAX}, the Fe K$\alpha$ line was  
almost invariably rather weak, with Equivalent 
Widths (EW) of a few tens of eV at most. Only in the observation  
of April 19, 1998, the line was strong enough 
(EW$\simgt 200$ eV; cp. Fig.~\ref{fig:line}) for a detailed  
analysis of its profile to be performed.   
We have therefore analyzed the spectral data from this observation, 
further selecting some intervals of relative calmness in  
the lightcurve (see Fig.~\ref{fig:1915lc}),  
to limit the spectral variability as much as possible. 
In order to model the source emission, we adopted the 
illumination model, local spectra and emissivities described 
in Martocchia, Matt \& Karas (2001) and Martocchia, Karas \& 
Matt (2000; see also: Martocchia 2000). Finally, we used
our own fitting tool, dedicated to GR spectral distortions in Kerr
metric, to analyze the data. The tool, named {\sc kerrspec},
is described in the Appendix. 

	\section{Data reduction} 
	\label{sec:data} 
 
{\it BeppoSAX} (Boella et al 1997) is composed by several  
instruments. Here we made use of the two imaging instruments,  
the LECS and the MECS, 
and the high energy, passively collimated instrument, the PDS. As 
customary, we have used the following energy bands: 0.1-4 keV 
for the LECS; 1.8-10 keV for the MECS; 15-150 keV for the PDS 
(neglecting higher energy data because of insufficient
statistics due to the short integration times).  
The LECS and MECS spectra have been selected from regions of 8' 
radius and centered on the source.  
The background has been evaluated from 
similar regions in blank field observations, and then subtracted. 
The PDS is composed by four units, two of 
which (switching every 96 s) monitoring the background, which is 
automatically subtracted in the pipeline producing the spectrum. 
 
We discuss here the {\it BeppoSAX} observation of GRS~1915+105 
started on April 19, 1998 at 11:33:19 UT and ended on April 20, 1998 
at 20:14:52 UT. 
Among the several {\it BeppoSAX} campaigns on GRS 1915+105, 
we selected this one because the iron line EW was significantly 
larger than in all other observations.

Out of the entire $\sim 120$ ks observation, 
some time intervals were sorted out where the source variability 
is less pronounced. These are the six satellite  
orbits starting from 3$\times$10$^4$~s 
after the beginning of the observation, see Fig.~1. Analysis of the other 
parts of the observation would require a finer temporal analysis which is 
beyond the scope of this paper and is deferred to a future work.  
 
All spectral fits were performed with the {\sc xspec} v.11 software package. 
In the following, all errors refer to 90\% confidence level for one  
interesting parameter ($\Delta\chi^2$=2.71) 
 
At the flux level of GRS~1915+105, it is very likely that systematic 
errors cannot be neglected with respect to the statistical ones. 
This may explain the rather large values of the $\chi^2$ reported below.  

\begin{figure*}[t]
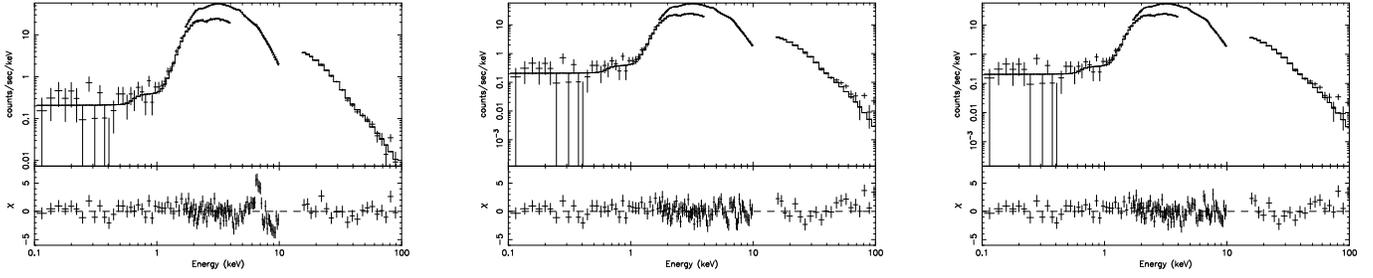
 
\includegraphics[angle=-90,width=0.3\textwidth]{h3276F2.ps} 
\hfill 
\includegraphics[angle=-90,width=0.3\textwidth]{h3276F3.ps} 
\hfill 
\includegraphics[angle=-90,width=0.3\textwidth]{h3276F4.ps} 
\caption{{\it Left:} The best-fit spectrum and  $\Delta\chi^2$ 
for the fifth interval, fitting without modelling the iron line  
($\chi^2_{\rm red}=2.64$): the residuals are evident. 
{\it Center:} The best-fit spectrum and $\Delta\chi^2$ 
for the same interval, when the line is modelled by a gaussian 
profile ($\chi^2_{\rm red}=1.48$).  
{\it Right:} The best-fit spectrum and $\Delta\chi^2$ 
for the fifth interval again, with a fully-GR model of the line profile 
($\chi^2_{\rm red}=1.31$). See the text for details. } 
\label{fig:models} 
\end{figure*} 

	\section{Data analysis and results} 
	\label{sec:datanal} 

	\subsection{The best-fit model in a selected time interval }

\begin{figure*}[t]
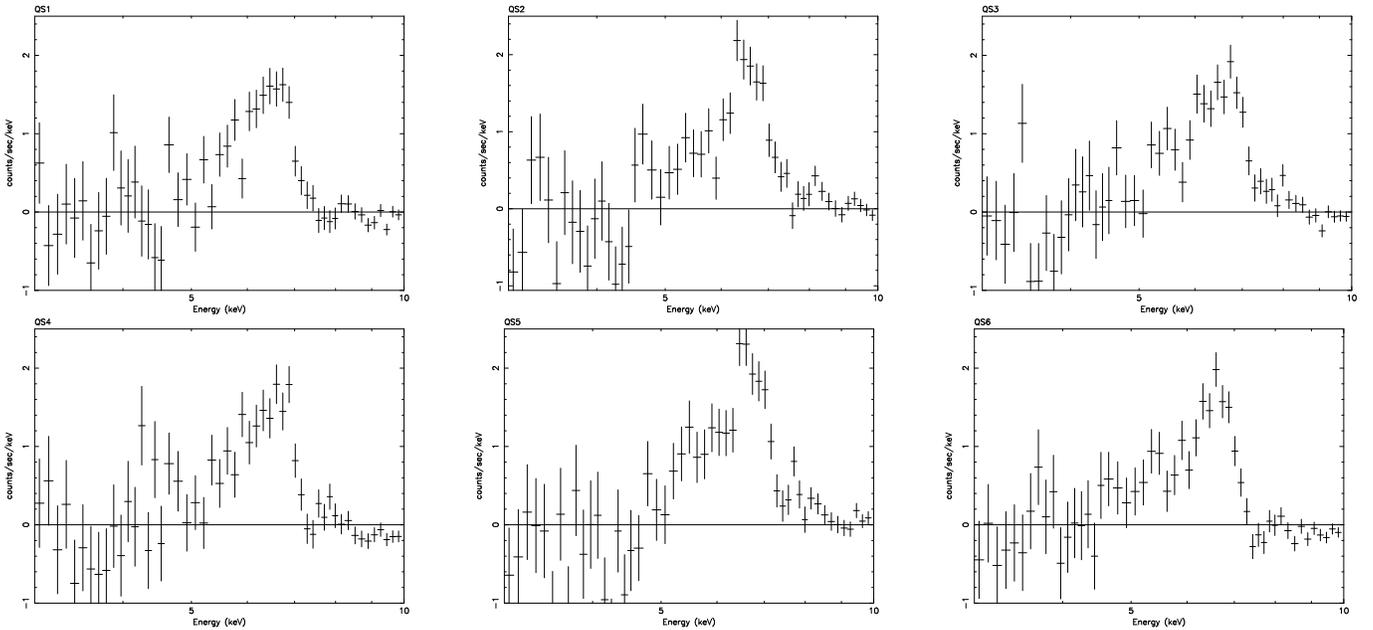
 
\includegraphics[angle=-90,width=0.3\textwidth]{h3276F5.ps} 
\hfill 
\includegraphics[angle=-90,width=0.3\textwidth]{h3276F6.ps} 
\hfill 
\includegraphics[angle=-90,width=0.3\textwidth]{h3276F7.ps} 
\hfill 
\includegraphics[angle=-90,width=0.3\textwidth]{h3276F8.ps} 
\hfill 
\includegraphics[angle=-90,width=0.3\textwidth]{h3276F9.ps} 
\hfill 
\includegraphics[angle=-90,width=0.3\textwidth]{h3276F10.ps} 
\hfill 
\caption{Residuals at the iron line energies for all the selected 
time intervals. For illustration purposes, the fits have been 
performed excluding the energy region of the iron line. Prominent,  
broad and skewed features are then evident around $\sim 6.4$ keV.} 
\label{fig:line}   
\end{figure*} 

Let us first discuss {\it the fifth interval} as an example, being the  
one in which a low value of  
$\chi^2$ was found when fitting with simple models.  
The model for the continuum includes the usual ingredients for this source: 
a disc thermal emission ({\sc diskbb}), a powerlaw with exponential cutoff, 
and a Compton Reflection component ({\sc pexrav}: Magdziarz \& Zdziarski 
1995), which is expected to always accompany the iron emission line.  
All these components are seen through interstellar absorption.  
The best-fit spectrum and  $\Delta\chi^2$ 
for the fifth interval, fitting without any iron line,  is shown in  
Fig.~\ref{fig:models}, left panel 
($\chi^2_{\rm red}=2.64$). The presence of an iron line is obvious. 
To better show the profile of the line,  
we also fitted the spectrum excluding the energy region which contains 
this feature (3-7.5 keV), and then reintroduced this energy interval. 
The residuals are shown in Fig.~\ref{fig:line} (fifth panel),  
after reinserting the data of the 3-7.5 keV interval. The presence 
of a prominent, broad and red-tailed iron line is clear.  
 
We first tried to model the line with a gaussian. 
The best--fit is obtained with a rest  
energy $E_0=6.38\pm0.12$ keV, $\sigma=0.70\pm0.15$ keV and EW$\sim250$ eV;  
the $\chi^2_{\rm red}$ is then 1.48 (see Fig.~\ref{fig:models}, central  
panel).
The most natural interpretation for a so broad line is  
emission from a relativistic accretion disc (Fabian et al., 1989), 
an hypothesis further supported by the asymmetry of the profile (see 
Fig.~\ref{fig:line}).  
The next step was therefore to substitute the gaussian 
with the relativistic model {\sc kerrspec}, described in the Appendix. 
To be self--consistent, we also applied the relativistic 
corrections to the reflection continuum (using {\sc kerrspec}) as well 
as to the disc thermal component (using {\sc diskpn}, Gierlinski et al.  
1999). After having verified, however, that the results are  
insensitive to the latter corrections, in order to save 
computer time we adopted the simple, non relativistic  
models for the continuum in all the following fits.  
The inclination angle has been always 
fixed to 70$^{\circ}$ (Mirabel \& Rodriguez 1994), 
while the inner and outer disc radii were kept free to vary.  
We tried different emissivity laws, and found the best solution  
with a rather ``flat'' dependence on radius ($\propto r^{-2}$). 
With the relativistic line model ($E_0=6.4$ keV)
the fit is better than with a gaussian: $\chi^2_{\rm red}=1.31$. 
The spectrum and residuals are  
shown in Fig.~\ref{fig:models}, right panel.  
If $E_0$ is instead fixed to 6.7 keV, corresponding to He--like iron, 
the fit is significantly worse ($\chi^2_{\rm red}>>2$). 

We also tried adding a narrow gaussian 
centered either at $6.4$ or at $6.7$ keV; no significant  
improvements in the fits were found.

One problem with the above fits is that 
the resulting parameters of the thermal emission are 
rather unplausible.  
The temperature is very large, $kT = 3.77^{+0.44}_{-0.43}$ keV,  
and the component's normalization very low ($0.41\pm 0.22$; 
see Arnaud 1996 for the meaning of this normalization), implying 
a value of the inner radius of the order of one  
kilometer. This, in turn, implies an upper limit to the  
BH mass (obtained assuming the inner emitting radius conciding with 
the innermost stable orbit for a maximally rotating BH)  
of less than 1 solar mass, contrary to observations (Greiner \ea, 2001b).
If the disc blackbody component is neglected in the model,  
the fit e.g. with the gaussian line yields a worst fit 
$\chi^2_{\rm red}=1.61$, but the results on the 
line parameters does not change 
significantly. Moreover, the fit with a relativistic line is 
still better than with a gaussian also in this case. 
 
\begin{table*} 
\centering 
\caption{The best-fit parameters for the various time intervals we 
considered, excluding LECS data and using a simplified model, 
which does not take into account thermal emission from the disc 
({\sc wabs ( pexrav + kerrspec )}). The rest line energy is 6.4 keV.
$\Delta{t}$ is the MECS exposure time. See the text for more details.} 
\vspace{0.05in} 
\begin{tabular}{|c|ccccccccc|} 
\hline 
& & & & & & & & & \cr 
\# interval & $\Delta{t}$ & $\Gamma$ & $E_{\rm c}$ & 
$R$ & $r_{\rm in}$ & $r_{\rm out}$ & 
EW & $F_{\rm 2-10}$ & $\chi^2_r$ \cr 
& & & & & & & & & \cr 
\hline 
& & & & & & & & & \cr 
& (s) & & (keV) & & ($\frac{GM}{c^2}$) & ($\frac{GM}{c^2}$) & (eV) &  
(10$^{-8}$~$\frac{{\rm erg}}{{\rm cm}^2{\rm s}}$) & (68 d.o.f.) \cr 
& & & & & & & & & \cr 
\hline 
\hline 
& & & & & & & & & \cr 
1 & 2562 & 2.638$_{-0.012}^{+0.013}$ & 25.6$_{-1.2}^{+2.6}$ & 
0.21$_{-0.20}^{+0.22}$ & 36$_{-17}^{+43}$ &  
330$_{-180}^{+320}$ & 142$^{+20}_{-10}$ & 1.876 & 1.27 \cr 
& & & & & & & & & \cr 
2 & 1993 & 2.702$_{-0.016}^{+0.017}$ & 26.7$_{-2.1}^{+2.4}$ & 
0.91$_{-0.31}^{+0.34}$ & 79$_{-34}^{+51}$ &  
710$_{-300}^{+610}$ & 192$^{+25}_{-30}$  & 1.776 & 1.78 \cr 
& & & & & & & & & \cr 
3 & 2551 & 2.653$_{-0.013}^{+0.013}$ & 28.2$_{-3.1}^{+2.7}$ & 
0.17$_{-0.17}^{+0.22}$ & 40$_{-24}^{+37}$ &  
350$_{-110}^{+170}$ & 192$^{+56}_{-23}$  
& 1.780 & 1.50 
\cr 
& & & & & & & & & \cr 
4 & 2003 & 2.666$_{-0.014}^{+0.013}$ & 28.8$_{-3.4}^{+5.0}$ & 
0.60$_{-0.09}^{+0.11}$ & 60$_{-31}^{+64}$ &  
180$_{-90}^{+190}$ & 156$^{+21}_{-19}$  & 1.799 & 2.58 \cr 
& & & & & & & & & \cr 
5 & 1628 & 2.675$_{-0.020}^{+0.023}$ & 23.8$_{-1.3}^{+3.9}$ & 
1.20$_{-0.41}^{+0.49}$ & 7.8$_{-3.7}^{+18.4}$  &  
240$_{-70}^{+100}$ & 298$^{+30}_{-39}$  
& 1.708 & 1.90 
\cr 
& & & & & & & & & \cr 
6 & 2708 & 2.608$_{-0.013}^{+0.014}$ & 23.6$_{-1.4}^{+5.4}$ &  
0.15$_{-0.15}^{+0.28}$ & 32$_{-13}^{+25}$ &  
210$_{-110}^{+170}$ & 149$^{+22}_{-22}$  & 1.863 & 1.75 \cr 
& & & & & & & & & \cr 
\hline 
\end{tabular} 
\label{newtable} 
\end{table*} 
 
We therefore conclude that the relativistic solution for the iron 
line, besides being the most natural physical explanation for the  
very broad line observed, does not depend on the details of the  
continuum model. 

The small relative luminosity of the disc thermal component (if present 
at all) brings us to interpret the  
source's state as a `C'-type, in the scheme of Belloni et al. (2000). 
This interpretation is in agreement with the occurrence of a  
``plateau'' in the light curve (lower flux and variability), 
although a hard spectrum would be expected, while instead  
we get a steep powerlaw, and a cutoff at rather  
low energies (see Table~1.).

	\subsection{Other time intervals }

For all other time intervals we found qualitatively similar  
results: a broad and prominent iron line, which is 
better fitted by a relativistic line profile
than by a gaussian, and a faint disc blackbody component.  
 
An inspection of the residuals often indicates  
problems both around 1 keV, and  
above 50 keV. The latter problem may be due to a non--thermal 
component arising when the Comptonized power law cuts--off,  
while we have no obvious explanation for the former problem.  
To overcome these difficulties, and because we are  
interested here to the iron line properties, we decided  
from now on to exclude the LECS data in the fits, and to  
use the PDS up to 50 keV only. With this prescriptions, 
and assuming as working hypothesis that the iron line arises from a 
relativistic disc, we analysed all the time intervals with a model  
which neglects disc thermal emission and includes, besides the 
line, a power law with exponential cut--off and a Compton reflection  
continuum. We stress again that the latter component is included  
more on astrophysical ground (it is supposed to always go along  
with the iron line) rather than because strongly required by  
the data. After having verified that the values of the  
absorbing column densities $n_{\rm H}$ for the various 
intervals are all consistent with each another and to the value of  
5.4$\times10^{22}$ cm$^{-2}$, for simplicity we fixed in all  
fits $n_{\rm H}$ to this value.  
 
The results for the various time intervals are summarized in  
Table~\ref{newtable}. The larger value of the $\chi^2_{\rm red}$  
of the fifth interval with respect to those described above  
is due to the lower number of spectral bins and to the exclusion  
of the thermal component. Given the abovementioned 
uncertainties on the proper modeling of the 
continuum, we stress that the values in the table 
should be taken with caution: a different continuum models may  
result in somewhat different values of the disc parameters.  
In any case, a general result seems to emerge: 
the inner radius of the line emitting region is larger than  
the innermost stable orbit. Whether this means that the  
innermost regions are absent, or simply that they do not emit  
the iron line (for instance because the matter is mildly  
ionized and resonant trapping effective, Matt et al. 1993) 
we cannot say.  
 
Finally, we tried summing the spectra of all six
time intervals together, and fitting with the same model. 
The very large reduced $\chi^2$, 5.1, indicates that  
spectral variability is too large over these time scales.  
In particular, strong residuals at the iron line energies  
imply that the line parameters change with time, 
as suggested by the results of the time--dependent spectral  
analysis. 
 
	\section{Discussion } 
 
A strong iron \ka\ fluorescent line is present in the April  
19, 1998, spectrum of the superluminal galactic source  
GRS 1915+105. The line is broad and skewed, clearly indicating 
emission from a relativistic accretion disc. We successfully  
fitted the line with a fully relativistic profile. To our 
knowledge, this is the first time that a relativistic iron line 
is observed in this source, largely due to the fact that we  
catched the line with an unusually large equivalent width.  

While providing evidence that an accretion disc 
resides in the innermost part of this system, our result
confirms that strong gravity is at work in distorting the 
iron line profile.
From the best--fit parameters we conclude that the line is from
neutral or low-ionization iron, and likely emitted from a region of  
the accretion disc which is close but probably does not  
include the innermost stable orbit, even
for a static BH (but the values of the disc parameters 
are somewhat dependent on the adopted model for the 
continuum, and must be taken with caution). This may 
imply either that the disc does not extend down to this  
radius, or that the line is not emitted in the innermost  
regions, maybe because the iron there is in the intervals 
of ionizations in which resonant trapping efficiently  
destroys line emission (e.g. Matt et al. 1993; 1996).  
Unfortunately, at these radii differences in the line  
profile between Kerr and Schwarzschild metrics  
are too subtle to be detectable, and the question of the 
angular momentum of the BH - which, from this analysis,
is not required to have a large value - actually remains
open. 

So, no final statement about the spin of the BH in
GRS 1915+105 can still be made. Zhang \ea\ (1997), on 
the basis of soft state X-ray brightness (which is most 
likely due to thermal emission), suggested that a fast 
rotating BH is hosted at the center of this source. 
Their conclusion was based on a mass value twice 
that measured by Greiner \ea\ (2001b). However, even 
accounting for the correct mass, the BH spin seems to 
be larger in GRS 1915+105 than in GRO J1655-40, and this 
does not fit to the results based on QPO diagnostics, 
as explained below. The accretion disc temperatures of 
the two sources are nearly identical, and the 
mass estimate for GRO J1655-40 is $M \sim 7 
M_\odot$ (Orosz \& Bailyn, 1997; Shahbaz \ea, 1999). Taking 
Zhang \ea's innermost thermally emitting orbit for GRS 1915+105
(at $\sim 40$ km, i.e. $r_{\rm ms} \sim 2r_{\rm g}$), one 
infers a BH spin $a \sim 0.95$ in this source (Bardeen \ea, 
1972). Actually, Zhang \ea's estimates of the BH angular momentum 
are based on rather simplified models of the continuum, and 
they do not include GR spectral distortions and effects of radiation 
transfer through the disc atmosphere (e.g. Merloni \ea, 
2000; Gierli\'nski \ea, 1999 and 2001). In our data analysis, 
even accounting for GR effects on the thermal component, 
we got disc temperatures 
and luminosities which are physically unplausible, and do not 
allow us to test assumptions regarding either the BH mass
nor its spin. Non-standard disc models (e.g. slim-disc) could
actually reproduce high temperature discs for non-rotating BHs.  

QPO diagnostics could therefore be useful as independent tests. 
Various models try to explain the occurrence of QPOs,  
suggesting that the oscillations correspond to the motion  
on the last stable orbit (Shapiro \& Teukolsky, 1983), and  
sometimes invoking GR effects such as ``frame dragging''  
and relativistic precession (e.g. Merloni \ea, 1999;  
Stella \ea, 1999). An interesting possibility is that the  
highest QPO frequency changes in connection with 
the variation of the disc innermost radius.  
 
QPOs in GRS 1915+105 have been observed in a wide range of 
frequencies ($0.001 \div 67$ Hz, e.g. Morgan \ea\ 1997).  
In GRS 1915+105 the QPO highest frequency (67 Hz),  
unlike all the others, does not change much with time,  
and may therefore be related to the  
accretion disc innermost stable orbit $r_{\rm ms}$  
(Morgan \ea\ 1997). A similar case is given by the  
300 Hz QPO in GRO J1655-40. 
 
Recently, a second high-frequency QPO has been detected 
in GRS 1915+105 (Strohmayer 2001), with $\nu_0 \sim 40$ Hz. 
In terms of the frame-dragging model, this lower  
frequency peak could be associated to the radial 
epicyclic frequency $\nu_{\rm rad}$ through the formula 
$\nu_0=\nu_{\rm Kep} - \nu_{\rm rad}$ (the nodal 
precession can be connected to a further low-frequency 
peak, already detected at $\sim 0.9$ Hz by Morgan \ea, 
1997). Strohmayer also reports that the 67 Hz QPO did in  
fact drift in frequency by as much as 5 percent.  
If this frequency is associated with the  
Keplerian motion at the last stable orbit, then - 
assuming the BH spin is the same in GRS 1915+105 and  
GRO J1655-40 - it should scale with $M_{\rm c}^{-1}$.  
Thus, a factor $\sim 2\div3$ of discrepancy would remain  
with respect to Greiner \ea's estimate. 
This indicates that the BH spin in the two sources is 
different. If the QPOs in GRS 1915+105 and GRO J1655-40 are seen 
in the framework of the frame dragging scheme (e.g. Cui \ea,  
1998, for details on the model), and using Greiner \ea's mass  
estimate, we have $a\sim 0.8$ and 0.95 for the two  
sources, respectively. On the other hand, if the QPO frequency 
had a diskoseismic origin (e.g. Nowak \ea, 1997), the model  
would require a maximally spinning BH in GRO J1655-40, and  
an almost static BH in GRS 1915+105 (cp. Greiner \ea, 2001b,
and references therein). 

Clearly, further tests for the BH spin would be of
extreme importance. Iron line
diagnostics may be crucial to solve the puzzle: 
better resolution and sensitivity spectral data,
provided by satellites of the new (e.g. {\it XMM-Newton}) and 
future (e.g. {\it Constellation-X}, {\it XEUS}) generations, will
be probably decisive in this respect.

\begin{acknowledgements} 
AM and GM acknowledge financial support from MURST under grant {\sc  
cofin-00-02-36}, GM also from ASI, VK from GAUK 188/2001 and GACR 
202/02/0735. TB thanks the Cariplo Foundation for financial support. 
\end{acknowledgements}

  \section*{Appendix: The computational technique }
  \label{sec:tool} 

We have computed GR effects  
acting on both the iron K$\alpha$ line emission and the Compton  
reflection continuum.  
We adopt the usual assumption of a geometrically thin, optically  
thick, cold (with respect to X-rays reprocessing)  
reflector extending over the equatorial plane. We take into account 
the directional anisotropy of the illuminating 
flux, caused by aberration, Doppler effect and light focusing.  
GR anisotropy of the illuminating flux is crucial to distinguish 
static from spinning BHs, because the primary 
illumination, as well as the induced emissivity of the 
reflection component, 
carry information about the spacetime in the  
very vicinity of the BH. The iron line profile and, in particular,  
its intensity and EW are affected by spacetime distortions
(Martocchia \& Matt, 1996; Martocchia \ea, 2000).  
 
The observed line profiles are often obtained  
from the data by subtracting the Compton-reflected 
component without taking into account important relativistic  
effects, such as the iron edge smearing, which may be relevant if the  
emission occurs down to the innermost stable orbit, in particular for 
spinning BH. In our computations (e.g.\ Martocchia \ea, 2000),  
besides using emissivity laws derived from a proper GR treatment  
of the illuminating flux, we treated at the same time  
the GR distortions on the iron line together  
with the underlying continuum hump 
(Martocchia \ea, 2000; Martocchia \ea, 2001). 

In order to fit real spectra, a ray-tracing code 
was developed in the form of a user-defined subroutine, {\sc kerrspec} 
(Martocchia 2000), which has been linked to the {\sc xspec} data 
analysis package (Arnaud 1996).
In the adopted scheme, a large set of light rays 
(null geodesics) in the Kerr BH spacetime is pre-calculated and 
stored. This allows an efficient access during the 
fitting loop. The code 
allows the radial, angular and energy dependencies of the local emissivity 
to be specified on input in a tabular or semi-analytical form. After 
fixing the desired grid resolution in the image plane, the required 
light rays corresponding to all image pixels are retrieved. 
 
The original approach (Karas \ea\ 1992) was refined in order to improve the 
resolution around the inner edge of the disc, very near to the BH 
horizon, and to ensure better confidence in the region where strong 
focusing occurs. This turned out to be necessary because the computed 
line profiles are very sensitive to the radius of the innermost emitting 
orbit, especially if the emissivity varies rapidly with radius. We 
recall that the innermost emission area is usually identified with a 
ring located at the marginally stable orbit, $r_1{\approx}r_{\rm{}ms}$, 
the value of which depending on the BH angular momentum. 
 
The local emissivity law, $F(r)$, must be specified before running the 
code. Different choices are available: 
a power-law dependence $F_q(r)\propto{r}^{-q}$; 
a ``standard'' Novikov-Thorne-Page model (Novikov \ea\ 1974);  
or the effect of illumination from a 
central source (Martocchia \& Matt 1996). 
In the latter case, $h$ (the primary source height) is a phenomenological 
parameter which determines the form of the local emissivity law: small $h$ 
corresponds to enhanced concentration of primary X-rays near the 
horizon, resulting in substantial anisotropy of the reflected component. 
The local emissivity has an angular dependence, too, 
including possibly a limb-darkening law. This may be expressed either by 
the functional forms for fluorescent and Compton-reflected emission 
given by Ghisellini \ea\ (1994), or, 
alternatively, by the numerically-derived tables from the 
Montecarlo computations by Matt \ea\ (1991), which yielded 
the whole spectra of the reflected component. 
 
The local profiles of the emissivity were interpolated by  
polynomials to save time machine (cp.\ Martocchia 2000, 
Martocchia \ea, 2001). These emissivity laws are valid between the 
inner disc edge (assumed to be equal to or greater than the innermost 
stable orbit) and the outer edge. Because the emission 
tends to be concentrated in the central parts of the disc, the inner 
emitting radius is usually better constrained 
in the fits than the outer radius. 
 
Typically, the code may take up to a few hours, on a 
common {\it DECalpha} workstation, to fit a spectrum  
using curved geodesics in the very vicinity of the BH, 
i.e. accounting for radiation emitted at distances of the order 
of $r_{\rm{g}}$ from the event horizon.

\end{document}